\documentclass[fleqn]{article}
\usepackage{emlines2,bezier,amscd}
\usepackage{amsmath}
\usepackage{amssymb}
\newcommand{\bt}{\ensuremath{\stackrel{b}{=}}}
\newcommand{\Dvg}{\ensuremath{{\rm Div}\,}}
\newcommand{\dvg}{\ensuremath{{\rm div}\,}}

\newcommand{\Gr}{\ensuremath{{\rm Grad}\,}}
\newcommand{\vl}[1]{\ensuremath{d{\rm vol}_{#1}}}

\newcommand{\frd}{\ensuremath{{}^{(4)}\!}}
\newcommand{\fvd}{\ensuremath{{}^{(5)}\!}}
\newcommand{\Rm}{\ensuremath{\mathrm{Riem}}}
\newcommand{\Rc}{\ensuremath{\mathrm{Ric}}}

\newcommand{\Li}[1]{\ensuremath{\pounds_{#1}}}

\title{NEMATIC STRUCTURE OF SPACE-TIME AND ITS TOPOLOGICAL DEFECTS
IN 5D KALUZA-KLEIN THEORY}
\author{Sergey S.Kokarev\thanks{Department of theoretical physics, r.409, YSPU,
                       Respublikanskaya 108, Yaroslavl, 150000, Russia, e-mail: sergey\symbol{"40}yspu.yar.ru}}
\date{}

\begin{document}
\maketitle

\begin{abstract}
\small
We show, that classical Kaluza-Klein theory possesses
hidden nematic dynamics. It appears as a consequence
of $1+4$-decomposition procedure, involving 4D observers
1-form $\lambda.$
After extracting of boundary terms the, so called, "effective matter"
part of 5D geometrical action becomes proportional to
square of anholonomicity
3-form $\lambda\wedge d\lambda$. It can be interpreted as twist nematic elastic energy, responsible
for elastic reaction of 5D space-time
on presence  of anholonomic 4D submanifold, defined by $\lambda.$
We derive both 5D covariant and $1+4$ forms of 5D nematic
equilibrium equations,
consider simple  examples and discuss some 4D physical aspects of
generic  5D nematic topological defects.
\end{abstract}

\vspace{0.2cm}  KEY WORDS: Kaluza-Klein theory, nematic structure,
anholonomic manifold\vspace{0.5cm}

\section{Introduction}\label{intro}

Up to a present time some fundamental concepts of continuum media mechanics have
revealed their relevancy for more deep understanding of space-time physics
\cite{born}-\cite{kok4}. As well as being evidence of interrelations and
unity of such, at first glance, remoted physical topics, this fact also
suggests that space-time and matter have unified geometro-physical
base, providing both physical interpreting of some subtle geometrical
properties of space-time and geometrical background for fundamental
properties of matter.

In present paper we turn our attention to a classical Kaluza-Klein theory
(KKT)\cite{kal,klein}. It attracts  many  theorists  today due
to profound insight of its central paradigm --- extradimensions and its physical
manifestations --- on the one hand, and due to   development
of the theory within more contemporary framework on the other
\cite{rum,wes1,pdl,kok5,others}.
The more general (than original Kaluza-Klein) formulations
and interpretings of the theory
have allowed  to establish unified geometrical background
for  a wide class of phenomena. Particularly,
problem of fifth force,
nature and origin of some fundamental classical notions (masses, charges),
actual  cosmological and astrophysical problems, some
important aspects of quantum mechanics and elementary
particle physics, especially, unification interactions problem --- all this  can be successfully
"translated" on the generalized
KKT language.

Let's remind that commonly used method of extracting
4D observable quantities from
5D world is 1+4-splitting procedure \cite{vlad,pdl}.
4D results of this procedure crucially depend on  particular choice
of 1-form $\lambda,$ providing local splitting of 5D riemannian
manifold on 4D space-time sections and extradirections.
This bring up the question: what  field $\lambda$ is realized for
a some 5D manifold with forgiven riemannian metrics $G$?
In previous works some authors  have considered
the freedom of choosing $\lambda$ as an effective instrument for obtaining
physically different 4D worlds from the same 5D manifold
(the so called "generational procedure" \cite{vlad,kok6}).
Other ones have restricted the freedom  using some additional
considerations (sometimes rather artificial).

In present paper we propose  natural framework
for answering to this question.
It is based on variational procedure,
applied to 4D part of 5D action.
In brief, let 5D geometrical  action\footnote{We assume $c=1$ and supply
all 4D and 5D quantities, denoted by the same letters with indexes
$(4)$ or $(5)$ when it is necessary.}
\begin{equation}\label{decact}
\fvd\mathcal{A}=-\frac{1}{2\kappa_5}\int\limits_{\mathcal{M}}
\fvd R,
\end{equation}
(where $\kappa_5,\ \fvd R$ --- 5D Einstein constant and  scalar curvature
respectively)
is decomposed as follows:
\begin{equation}\label{decsymb}
\fvd\mathcal{A}=\frd\mathcal{A}[\lambda]+
\frd\mathcal{A}_{\rm m}[\lambda].
\end{equation}
Here $\frd\mathcal{A}[\lambda]$ ---   action for 4D gravity  and
$\frd\mathcal{A}_{\rm m}[\lambda]$ --- action for 4D
effective matter. If 5D metrics $G$, minimizing $\fvd\mathcal{A},$
is fixed,
then $\fvd\mathcal{A}$ does not depend on $\lambda,$
while  the both terms in the righthand
part of (\ref{decsymb}) are depend (this is reflected in notations).
We can assume, that {\it true  $\lambda$ is
extremal for  any of this two terms} (if one maximal, then
other minimal and vice versa).
For the definiteness we'll take
$\delta\frd\mathcal{A}_{\rm  m}[\lambda]=0$
as equations, determining $\lambda.$
At this point we'll reveal remarkable analogy of the problem to
similar 3D problem for  equilibrium deformations
of nematic liquid crystals in continuum media physics \cite{land}.
We'll see that unit 4D observers field $\jmath_\lambda,$ dual to
$\lambda,$ plays role of director
and endows 5D space-time nematic structure.

The plan of the paper is as follows.

In Sec. \ref{1+4} we remind some basic ideas,
relations and expressions of 5D KKT in frame of
$1+4$-splitting formalism. The aim of Sec.\ref{1+4}
is expression
(\ref{m4}) for  lagrangian (up to the  constant $-1/2\kappa_5$)
for action $\frd\mathcal{A}_{\rm m}$
in (\ref{decsymb}).

Sec. \ref{var1} is devoted to nematic structure
of 5D space-time, inspired by action $\frd\mathcal{A}_{\rm m}$
for effective matter. We remind general theoretical
assumptions of nematic liquid crystal physics.
Then we rewrite geometrical 4D lagrangian in the form, which
clears analogy with nematic crystals and
compare action
$\frd\mathcal{A}_{\rm m}$ with elastic  nematic free energy.

In Sec. \ref{vp}  we derive equilibrium equations  from
action $\frd\mathcal{A}_{\rm m}$ together with
boundary conditions.

In Sec.\ref{obs} $1+4-$form of 5D nematic equilibrium equations is presented.

Small Sec. \ref{part} touches some particular solutions, which can be
observed in earlier works.

In Sec. \ref{ex}  some  examples of 5D nematic structures,
satisfying equilibrium equations, are performed.

Conclusion contains general discussion.

In mathematical notations we follow mainly to \cite{neil,griffits}.
Particularly, we use the following notations and abbreviations:

$\imath_X$ and $\jmath_\omega$ --- 1-form and vector field dual to
vector field $X$ and 1-form $\omega$ respectively.
For any vector field $Y$  we have:
$\imath_X(Y)\equiv\langle X,Y\rangle,$ $\langle\jmath_\omega,Y\rangle\equiv
\omega(Y),$ where $\langle\ ,\ \rangle$ --- riemannian metric;

$T^r_s(\mathcal{M})$ --- $r$-contravariant and $s$-covariant tensor fields
over $\mathcal{M}$;

$\mathcal{T}(\mathcal{M})=\bigoplus\limits_{r,s} T^r_s(\mathcal{M})$ --- tensor algebra over $\mathcal{M}$;

For any $T\in T^0_2(\mathcal{M})$ we define $\jmath T$ and $T\jmath$
by the formulae:
\[
(\jmath T)(\omega, X)=T(\jmath_\omega,X);\quad (T\jmath)(X,\omega)=T(X,\jmath_\omega).
\]
Coordinate form: $(\jmath T)^\alpha_\beta=G^{\alpha\gamma}T_{\gamma\beta},$
$(T \jmath)^\alpha_\beta=G^{\alpha\gamma}T_{\beta\gamma}$ shows,
that $\jmath$  can be viewed as coordinate free notation
of tensor indexes
raising. Lowering is defined similarly by means of $\imath;$

$\hat{\mathcal{S}}$ and $\hat{\mathcal{A}}$ --- symmetrization and
antisymmetrization operators, acting in spaces $T^n_0(\mathcal{M})$ and
$T_n^0\mathcal{M}$ for every $n$; for example, in case $T\in
T^0_2(\mathcal{M}):$
\[
(\hat{\mathcal{S}}T)(X,Y)=\frac{1}{2}(T(X,Y)+T(Y,X));\quad
(\hat{\mathcal{A}}T)(X,Y)=\frac{1}{2}(T(X,Y)-T(Y,X));
\]

$\nabla:\ T^r_s(\mathcal{M})\to T^r_{s+1}(\mathcal{M})$ covariant
(with respect to some fixed riemannian metrics $G$) derivative;

$\Gr\equiv\jmath_{\nabla}$; $\Dvg X={\rm Tr}(\nabla X)$; $\Dvg\omega=
{\rm Tr}(\Gr\omega)$ --- some useful differential operations, connected
with $\nabla.$ Here $X$ and $\omega$ ---  arbitrary vector field and 1-form.

$\pi_X:\ \Lambda^p(\mathcal{M})\to\Lambda^{p-1}(\mathcal{M})$ --- lowering
degree operator, acting on space of external forms of degree $p$ by the rule:
\[
(\pi_X\omega)(Y_1,\dots,Y_{p-1})=\omega(X,Y_1,\dots,Y_{p-1});
\]

$\vl{5}\equiv\sqrt{|G|}dx^0\wedge dx^1\wedge dx^2 \wedge dx^3\wedge dx^5$ ---
standard volume form on 5D riemannian manifold $\mathcal{M}$ with metric
$G.$

\section{Essentials of $1+4-$approach to 5D KKT}\label{1+4}

Present section is brief resemblance of some general ideas and relations
of $1+4$-splitting technic (monad method). In what follows
we'll use coordinate free formalism
and  modern apparatus of differential geometry. General scheme of
the section
is similar to \cite{vlad}, where one also can  find some details
in classical coordinate form.

\subsection{Algebra of monads method}\label{algebra}

Let  $\mathcal{M}$ be (semi-)riemannian 5-dimensional
manifold\footnote{In this section our consideration will be local, so
we don't fix global topology on $\mathcal{M}$.}
with some fixed metric $G\equiv\langle\ ,\ \rangle.$
The most general way to go from the 5D world to some "embedded" 4D is to fix
smooth
{\it 4D observers 1-form} $\lambda$ ({\it monad field}):
\begin{equation}\label{norm}
\langle\lambda,\lambda\rangle=\varepsilon,
\end{equation}
where we leave freedom of causal type of  direction $\jmath_\lambda$
by means of constant factor $\varepsilon=\pm1.$
The form induces decompositions of
tangent and cotangent spaces at every point $p\in\mathcal{M}$:
\begin{equation}\label{decomp}
T_p\mathcal{M}=(T_p)_h\mathcal{M}\oplus \jmath_\lambda(p);\ \ \
T_p^\ast\mathcal{M}=(T_p^\ast)_h\mathcal{M}\oplus \lambda(p),
\end{equation}
where {\it horizontal} tangent and cotangent spaces are:
\[
(T_p)_h\mathcal{M}\equiv\{v\in T_p\mathcal{M}\,|\, \lambda(v)_p=0\}\quad\mbox{\rm and}\quad
(T_p^\ast)_h\mathcal{M}\equiv\{\omega\in T^\ast_p\mathcal{M}\,|\, \omega(\jmath_\lambda)_p=0\}
\]
respectively. The subspaces ${\rm span}_p(\jmath_\lambda)$ and
${\rm span}_p(\lambda)$ we'll call {\it vertical.}
Let's note, that the set
\[
T_h\mathcal{M}\equiv\bigcup\limits_{p\in\mathcal{M}}(T_p)_h\mathcal{M}
\]
(or  similarly  $T^\ast_h\mathcal{M}$) in general does'nt admit local
representation $R\times T(\mathcal{M}_h)$ where   $\mathcal{M}_h$ --- {\it
horizontal manifold,} since form $\lambda$ can be {\it anholonomic (nonintegrable)}.
In this
situation we'll refer to $\mathcal{M}_h$ as
{\it anholonomic horisontal manifold}
\cite{griffits}, such that formally\footnote{In case of the, so called, {\it complete
nonintegrability,} Rashevski-Chow's  theorem \cite{rash,chow} states, that $\mathcal{M}_h=\mathcal{M}$
i.e. any two points of $\mathcal{M}$ can be joined by a some horisontal
curve $\gamma_h.$} $T(\mathcal{M}_h)\equiv T_h\mathcal{M},$
keeping in mind that just the case is realized, when we observe 4D world filled
by  electromagnetic fields from the viewpoint of 5D KKT (see Sec.\ref{lagr}
below).

Tensor continuations of (\ref{decomp}) give decomposition of a whole tensor
algebra $\mathcal{T}(\mathcal{M})$ on   $\lambda-h$ components. Formally,
let consider linear operator (affinnor field):
\begin{equation}\label{aff}
\hat h\equiv \hat I-\varepsilon \lambda\otimes \jmath_\lambda\equiv\hat I-\hat\lambda,
\end{equation}
mapping $T\mathcal{M}\to T\mathcal{M}$ and
$T^\ast\mathcal{M}\to T^\ast\mathcal{M}.$ Here
$\hat I={\rm id}_{T\mathcal{M}}$ or $\hat I={\rm id}_{T^\ast\mathcal{M}}.$
By the definition it follows, that
$\hat h(\hat h(X))=\hat h(X)$  and $\langle\hat h(X),Y\rangle=0$
for every vector field $X$ and every vertical $Y$
(the same is true for 1-forms).
So, $\hat h$ is projector:
$T\mathcal{M}\stackrel{\hat h}{\to} T_h\mathcal{M}$ or
$T^\ast\mathcal{M}\stackrel{\hat h}{\to} T^\ast_h\mathcal{M}.$
Writing $\hat I=\hat \lambda+\hat h$
and taking its $n$-th tensor degree, we have:
\begin{equation}\label{decompun}
\hat I^{\otimes n}\equiv{\rm id}_{T^r_{n-r}(\mathcal{M})}=(\hat\lambda+\hat h)^{\otimes n}
=\sum\limits_{\varsigma}\hat\pi_\varsigma,
\end{equation}
where $\varsigma$ runs all binary sequences of symbols $\{\lambda,h\}$
of length $n,$
$\hat\pi_\varsigma$ --- projector on $\varsigma-$th component of
$T^r_{n-r}(\mathcal{M}).$
Acting by initial  and final operators of
(\ref{decompun}) on any tensor field
$T\in T^r_{n-r}(\mathcal{M})$, we have
\begin{equation}\label{decompt}
T=\sum\limits_{\varsigma}T_\varsigma,
\end{equation}
where  $T_\varsigma=\hat\pi_\varsigma(T)$ --- $\varsigma$-th
projection of $T.$ In what follows we'll denote projections by index-like
symbols $\lambda$ or $h$  when it will not lead to ambiguousness.
For example, any vector field can be decomposed as follows:
$X=\varepsilon X_\lambda \jmath_\lambda+X_h,$
where $X_\lambda\equiv\lambda(X),$ $X_h\equiv\hat h(X).$

With using (\ref{decompt}) it is easy to get  decomposition of $G$:
\begin{equation}\label{dec}
G=\varepsilon\lambda\otimes\lambda+h,
\end{equation}
where $h$ is {\it metric on (anholonomic) manifold}  $\mathcal{M}_h,$
defined by the rule:
\begin{equation}\label{hprod}
h(X,Y)=\langle\hat h(X),\hat h(Y)\rangle
\end{equation}
for any vector fields $X,Y.$
(\ref{hprod}) means, that
$h(X,Y)=G(X,Y)$
for every horizontal vector fields $X=X_h$ and $Y=Y_h$ and
$
\ker\,h={\rm span}(\jmath_\lambda).
$

Physically, any $\lambda$ defines smooth family of 4D observers histories,
which trace out 4D worlds inside the given 5D world. Accordingly to some modern
concepts, suggested by brane physics and quantum mechanics,   1-form
$\lambda$ should be related to perceptive
spaces of an observers \cite{kok7}, that we'll discuss in Conclusion.

\subsection{1+4-analysis on $\mathcal{M}.$}\label{analysis}

By (\ref{norm}) it follows, that\footnote{Here and below  we use abbreviated notation
$D_\omega\equiv D_{\jmath_\omega}$
for any kind of derivative $D$ along vector field  $\jmath-$conjugated with some $1-$form
$\omega.$} $(\nabla_\lambda\lambda)_\lambda=0.$
Let define {\it horizontal curvature 1-form of  $\jmath_\lambda-$congruence}:
\[
\alpha\equiv\nabla_\lambda\lambda.
\]
It is obviously,  that the tensor
$
\mathcal{H}\equiv\nabla\lambda-\varepsilon\lambda\otimes\alpha
$
is horizontal. It can be decomposed on  symmetric and
antisymmetric components:
$\mathcal{H}=\mathcal{D}+\mathcal{F},$
where
\[\mathcal{D}\equiv\hat{\mathcal{S}}(\nabla\lambda-\varepsilon\lambda\otimes\alpha)=
\frac{1}{2}(\pounds_\lambda G-\varepsilon\lambda\vee\alpha)
\]
--- {\it 4D extrinsic curvature tensor,}
\[\mathcal{F}\equiv\hat{\mathcal{A}}(\nabla\lambda-\varepsilon\lambda\otimes\alpha)=
\frac{1}{2}(d\lambda-\varepsilon\lambda\wedge\alpha)\]
--- {\it 4D twist tensor}
and $a\vee b\equiv a\otimes b+b\otimes a.$
Finally, we obtain:
\begin{equation}\label{decf}
\nabla\lambda=\varepsilon\lambda\otimes\alpha+\mathcal{H}.
\end{equation}
Acting in (\ref{decf}) by $\jmath$ from the right (with using $[\nabla,\jmath]=0$),
we obtain for vector
field $\jmath_\lambda$:
\begin{equation}\label{decv}
\nabla \jmath_\lambda=\varepsilon\lambda\otimes \jmath_\alpha+\mathcal{H}\jmath.
\end{equation}
Following to \cite{vlad}, let define operators of {\it vertical}
and {\it horizontal} (4D space-time) {\it derivatives}:
\[\dot T_h\equiv\frac{d}{d\lambda} T_h \equiv(\pounds_\lambda T_h)_h;\ \ \ {}^{(4)}\nabla T_h\equiv(\nabla_h T_h)_h,\]
where $T_h$ --- arbitrary horizontal tensor field. On scalar functions by definition:
\[
\dot f\equiv \jmath_\lambda(f);\ \ \ {}^{(4)}\nabla f\equiv (df)_h\equiv d_hf.
\]
With using (\ref{decompt}) the following identity for any vector field $Z$
can be established:
\begin{equation}\label{deccov}
\nabla Z={}^{(4)}\nabla Z_h+\varepsilon Z_\lambda\mathcal{H}\jmath+
(\dot Z_\lambda-Z_\alpha)\lambda\otimes \jmath_\lambda+
\lambda\otimes(Z_\lambda \jmath_\alpha+\varepsilon( \dot Z_h+\mathcal{H}\jmath(Z_h,\ ))
\end{equation}
\[
+\varepsilon((dZ_\lambda)_h-\mathcal{H}(\ ,Z_h))\otimes \jmath_\lambda,
\]
where $Z_\alpha=\alpha(Z).$
Acting on (\ref{deccov}) by $\imath$ from the right,
identifying $\imath_Z\equiv\omega$ and using
the relation
\[
\imath_{\dot Z_h}=\frac{d}{d\lambda}\imath_{Z_h}-2\mathcal{D}(Z_h,\ ),
\]
 we have for 1-forms:
\begin{equation}\label{deccovf}
\nabla\omega={}^{(4)}\nabla \omega_h+\varepsilon \omega_\lambda\mathcal{H}+
(\dot \omega_\lambda-\omega_\alpha)\lambda\otimes\lambda+
\lambda\otimes(\omega_\lambda \alpha+\varepsilon( \dot \omega_h-\mathcal{H}\jmath(\ ,\omega_h ))
\end{equation}
\[
+\varepsilon((d\omega_\lambda)_h-\mathcal{H}\jmath(\ ,\omega_h))\otimes \lambda.
\]
Assuming in (\ref{deccovf}) $\omega=\lambda,$ $\omega_\lambda=\varepsilon,$
$\omega_h=\omega_\alpha=0$ we obtain (\ref{decf}).

The formulae (\ref{deccov})-(\ref{deccovf}) show, that any
5D expression, including   covariant derivatives can be
reexpressed in terms of vertical and horizontal derivatives.
The following useful identities are easy checked:
\[
\dot\lambda=\alpha=\Li{\lambda}\lambda;\ \ \ {}^{(4)}\nabla\lambda\equiv(\nabla_h\lambda)_h=\mathcal H;\ \ \ \dot h=2\mathcal{D}=\dot G  ;\ \ \ {}^{(4)}\nabla h=0.
\]
The latter expression suggests, that operator ${}^{(4)}\nabla$ should be
treated as
"covariant"\footnote{In fact, $\frd\nabla$ possesses effective
torsion, since direct
calculation gives:
${\rm Tors}_{\frd\nabla}(X_h,Y_h)\equiv\frd\nabla_{X_h}Y_h-\frd\nabla_{Y_h}X_h-
[X_h,Y_h]=2\varepsilon\mathcal{F}(X_h,Y_h)\jmath_\lambda.$
However, with respect to {\it horisontal bracket:} $[\cdot{}_h,\cdot{}_h]_h$
torsion of $\frd\nabla$ is zero.}
(relatively $h$) derivative on $\mathcal{M}_h.$

\subsection{Effective matter lagrangian  in $1+4-$formalism}
\label{lagr}

Twice substituting (\ref{deccov})
into the definition of curvature operator:
\[
\Rm(X,Y)Z\equiv(\nabla_Y\nabla_X-\nabla_X\nabla_Y+\nabla_{[X,Y]})Z
\]
and twice appropriately contracting the obtained expression,
after some  $1+4$ algebra, outlined in previous subsection, we obtain:
$\fvd R=\frd R+M,$
where   $\frd R$ --- {\it 4D scalar curvature}
and
\begin{equation}\label{m4}
M=
2\alpha^2-2\varepsilon{\rm div}\,
\jmath_\alpha+
2\varepsilon\dot{\overline{\mathcal{D}}}+
\varepsilon(\overline{\mathcal{D}}^2+\mathcal{D}^2)
+\varepsilon\mathcal{F}^2
\end{equation}
--- {\it matter scalar}\footnote{The expression (\ref{m4}) corresponds to the formula (11.20) on p.218 in \cite{vlad}.
The correspondence is established with taking into account that: 1)
author works in special (the, so called, chronometric) gauge of monad
formalism, that requires fixing of coordinate system adopted to $\jmath_\lambda;$
2) curvature operator in \cite{vlad} has opposite sign to
accepted in present work; 3) $-\varepsilon\Phi$ in \cite{vlad} corresponds
to our $\alpha,$ $-\varepsilon\widetilde{F}$ corresponds to $\mathcal{F}.$
Our $\mathcal{D}$ is the same as in \cite{vlad}.}. Here $T^2\equiv\langle T,T\rangle$
for any tensor field $T,$ $\dvg X_h\equiv {\rm Tr}(\frd\nabla X_h),$
$\overline{T}\equiv {\rm Tr}(T\jmath)$ for any $T\in T^0_2(\mathcal{M}).$

So, 5D KKT inspires the following action for $\lambda$:
\begin{equation}\label{act4}
\mathcal{A}_{\text{m}}[\lambda]=-\frac{1}{2\varkappa_5}
\int\limits_{\mathcal{M}} M\,\vl{5}
\end{equation}
with $M$ given by (\ref{m4}). (\ref{act4}) is starting point for
our following consideration.

\section{Nematic structure of 5D space-time}\label{var1}

Since now we are interested  by extreme form $\lambda$
on $\mathcal{M}$ with forgiven vacuum metric $G,$
it would be  more appropriately
temporary to go aside  from the standard
view in KKT and  rewrite (\ref{m4})
in terms of  $\lambda$ and its 5D covariant
derivatives\footnote{Direct calculations with (\ref{m4}) are also
possible, but
take much more efforts, since 4D and $\lambda$-derivatives don't commute
with variations $\delta,$ while for example $[\nabla,\delta]=0$
since $G$ is fixed.}.
After little algebra (\ref{m4}) can be performed as follows:
\begin{equation}\label{lform}
M=-(\nabla_\lambda\lambda)^2+
2\varepsilon\jmath_\lambda(\Dvg\jmath_\lambda)+
\varepsilon(\Dvg\jmath_\lambda)^2-2\varepsilon\Dvg
\nabla_\lambda\jmath_\lambda
+\varepsilon(\nabla\lambda)^2.
\end{equation}
Before deriving equilibrium equations let's clear out nematic
properties of space-time with lagrangian (\ref{lform}).
For this purpose we need to remind general facts of
common 3D nematic crystals physics and generalize it for 5D case.
Nematic liquids form a subclass of fluid bodies with homogeneous but
anisotropic correlation function,  possessing axial symmetry \cite{land1}.
In other words, at every point of nematic liquid there is direction, connected
with   orientation anisotropy of single molecules, which
the liquid consist of.
Macroscopically this situation  can be described
by means of unit vector field $n,$ named {\it field of director}.
It should  be, in fact,
understood as an element  of unit projectivized tangent bundle
$U_P\approx R^3\times RP^2$,
since  directions $n$ and $-n$ for nematic are physically  equivalent.
Absolute nematic energy minimum  is realized  under $n={\rm const},$
while nonuniform field   $n$ describes possible deformed state of nematic.
Elastic (free) energy density of such deformed state within linear theory
can be expressed through invariant quadratic combinations of derivatives
$\nabla n$ possessing all required symmetry properties.
Up to a boundary terms nematic elastic energy density has  the following general kind
\cite{land1}\footnote{\label{mist} There is misprint in this book in expression
for invariant $(\nabla n)^2$ in \S 140. Right (and exact)
expression is
$(\nabla n)^2=(n,{\rm rot}\, n)^2+({\rm div}\, n)^2+(\nabla_nn)^2
+{\rm div}(\nabla_nn-n\,{\rm div}\, n).$}:
\begin{equation}\label{nemen}
F=\frac{K_1}{2}({\rm div}\, n)^2+\frac{K_2}{2}(n,{\rm rot}\, n)^2+
\frac{K_3}{2}(\nabla_n n)^2,
\end{equation}
where $K_1,K_2,K_3$ --- {\it Frank's moduluses}, responsible for {\it splay,
twist} and {\it bend} nematic elasticity respectively
(here temporary $\dvg$ means
3D divergency, ${\rm rot}$ --- standard 3D curl, $\nabla$ --- covariant derivative
in 3D euclidian space, (\ ,\ ) --- 3D euclidian scalar product). Some interesting problems,
concerning static nematic deformations
and their topological properties, whose 5D analogies  we'll consider lately,
can be found in \cite{land} (\S 36-39).

Easy to see, that 5D space-time in KKT
can be treated as some nematic medium in the problem of extreme $\lambda$
finding, since the lagrangian $L_\lambda=-(1/2\kappa_5)M$ with
$M$ given by (\ref{lform}) has  similar to (\ref{nemen}) structure.
This remarkable analogy  suggests
once again that space-time (4D or multidimensional)
can manifest properties of continuum media in various aspects, --- the fact,
that make such analogies useful for studying, interpreting and modeling
of space-time physics.

To express (\ref{lform}) in terms of Frank's moduluses we need express
it in terms of  independent quadratic
invariant combinations, which are 5D
generalizations of those in (\ref{nemen}). The
first and second terms in (\ref{nemen}) have trivial generalizations:
\[
({\rm div}\, n)^2\to (\Dvg\jmath_\lambda)^2; \ \
(\nabla_n n)^2\to (\nabla_\lambda\lambda)^2.
\]
The expression $(n,{\rm rot}\, n)^2,$ which is called  {\it anholonomicity of 3D vector field $n$},
has direct generalization $(\lambda\wedge d\lambda)^2/3!,$
since both expressions guarantee local integrability of 1-forms $\imath_n$ and $\lambda$
respectively and in 3D euclidian space second is identical to the first.
Using   relations:
\[
X(f)\bt-f\Dvg X;\ \  (\lambda\wedge d\lambda)^2\bt
3!(\varepsilon((\nabla\lambda)^2-(\Dvg\jmath_\lambda)^2)-
(\nabla_\lambda\lambda)^2)
\]
where $X,f$ --- arbitrary vector field and scalar function respectively,
"$\bt$" means "is equal up to a total divergence"
(the second identity in 3D space  under $\varepsilon=1$
is the formula of footnote \ref{mist}
up to a boundary terms), we obtain from (\ref{lform}):
\begin{equation}\label{lagnem}
\mathcal{L}_\lambda\bt-\frac{1}{12\kappa_5}(\lambda\wedge d\lambda)^2.
\end{equation}

From (\ref{lagnem}) we see that:
\begin{enumerate}
\item
{\it Nematic elasticity of 5D space-time, inspired by 5D KKT,
concerns only  nematic twists;}
\item
{\it Nonzero Frank's modulus $K_2=-1/6\kappa_5$ is induced
by 5D gravity.} The similar relation between Young's modulus
of multidimensional space-time and 4D Einstein constant $\varkappa$
has been observed in \cite{kok1} in the context of
"common" elasticity of space-time;
\item
From the view point of 4D physics  {\it 5D nematic elasticity
characterizes resistance   of 5D space-time with respect to anholonomicity
of embedded 4D physical worlds.} In other words, 5D "nematic vacuum"
contains only  holonomic physical world(s), traced out by 1-form
$\lambda$ with $\lambda\wedge d\lambda=0.$
\end{enumerate}

\section{Variational problem}\label{vp}

Varying modified action (\ref{act4}) with lagrangian (\ref{lform})\footnote{We go back from lagrangian
(\ref{lagnem}) to (\ref{lform}) in order to
obtain right boundary conditions.}:
\[
\mathcal{A}[\lambda]=-\frac{1}{2\kappa_5}\int\limits_{\mathcal{M}}(M -Q[\lambda^2-\varepsilon])\,\vl{5},
\]
including Lagrange multiplier $Q,$ after standard extracting
of exact forms we obtain the following volume part of variation:
\begin{equation}\label{volvar}
\delta\mathcal{A}_{\rm vol}=
-\frac{1}{\kappa_5}\int\limits_{\mathcal{M}}\delta\lambda\left[
\nabla^2_\lambda\jmath_\lambda-
\langle{\rm Grad}\,\lambda,\nabla_\lambda\lambda\rangle
+\Dvg\jmath_\lambda\,\nabla_\lambda\jmath_\lambda+
\varepsilon({\rm Grad}\,\Dvg\jmath_\lambda-\nabla^2\jmath_\lambda)-Q\jmath_\lambda
\right]\, \vl{5}
\end{equation}
and boundary terms
\[
\delta\mathcal{A}_{\rm b}=-\frac{1}{\kappa_5}\int\limits_{\partial\mathcal{M}}
\left[(\varepsilon\Dvg\delta\jmath_\lambda-\langle\delta\lambda,\nabla_\lambda\lambda\rangle)\,
\pi_{\jmath_\lambda}-\varepsilon(\pi_{\nabla_{\delta\lambda}\jmath_\lambda}+
\pi_{\nabla_\lambda\delta\jmath_\lambda})+\varepsilon\pi_{\jmath_{(\nabla\lambda)(\ ,\delta\jmath_\lambda)}}
\right]\, \vl{5},
\]

By arbitrariness of $\delta\lambda$   (\ref{volvar})  gives
the following 5D covariant nematic equilibrium
equations:
\begin{equation}\label{eqgen}
\nabla^2_\lambda\jmath_\lambda-
\langle{\rm Grad}\,\lambda,\nabla_\lambda\lambda\rangle
+\Dvg\jmath_\lambda\,\nabla_\lambda\jmath_\lambda+
\varepsilon({\rm Grad}\,\Dvg\jmath_\lambda-\nabla^2\jmath_\lambda)-Q\jmath_\lambda=0
\end{equation}
Its $\lambda$-component defines Lagrange multiplier $Q$:
\[
\varepsilon Q=(\nabla^2_\lambda\jmath_\lambda)_\lambda-
(\nabla_\lambda\lambda)^2
+\varepsilon(\nabla_\lambda\Dvg\jmath_\lambda-(\nabla^2\jmath_\lambda)_\lambda).
\]
Physical meaning has
$h$-component of (\ref{eqgen}):
\begin{equation}\label{eqh}
(\nabla^2_\lambda\jmath_\lambda)_h-
\langle{\rm Grad}_h\lambda,\nabla_\lambda\lambda\rangle
+\Dvg\jmath_\lambda\,\nabla_\lambda\jmath_\lambda+
\varepsilon({\rm Grad}_h\,\Dvg\jmath_\lambda-(\nabla^2\jmath_\lambda)_h)=0.
\end{equation}

\section{1+4-form of nematic equations}\label{obs}

For interpreting
of equations (\ref{eqh}) it is more
convenient to go again to $1+4$-representation.
Using the identities:
\[
(\nabla^2_\lambda\jmath_\lambda)_h=\dot\jmath_\alpha+\mathcal{H}\jmath(\jmath_\alpha,\ );
\quad \langle{\rm Grad}_h\lambda,\nabla_\lambda\lambda\rangle=\jmath\mathcal{H}(\ ,\jmath_\alpha);\quad
\Dvg\jmath_\lambda\,\nabla_\lambda\jmath_\lambda=\overline{\mathcal{D}}\jmath_\alpha;
\]
\[
{\rm Grad}_h\Dvg\jmath_\lambda=d_h\overline{\mathcal{D}};\quad
(\nabla^2\jmath_\lambda)_h=\varepsilon\overline{\mathcal{D}}\jmath_\alpha+
\varepsilon\dot\jmath_\alpha+\varepsilon\mathcal{H}\jmath(\jmath_\alpha,\ )+
\dvg\mathcal{H},
\]
equations (\ref{eqh}) can be rewritten in the
following  equivalent $1+4$ form:
\begin{equation}\label{eqh14}
\varepsilon\dvg\mathcal{F}\jmath=
\varepsilon {\rm grad}\,\overline{\mathcal{D}}-
\varepsilon{\rm div}\,\mathcal{D}\jmath
-\jmath\mathcal{H}(\ ,\jmath_\alpha),
\end{equation}
where ${\rm div}\, T_h=\overline{\frd\nabla\jmath}T_h,$ ${\rm grad}\equiv\jmath_{\frd\nabla}.$
In such form, it can  be interpreted as follows:
{\it origins of twists of $\mathcal{M}_h$}
(and consequently anholonomicity, since $\lambda\wedge d\lambda=2\lambda\wedge\mathcal{F}$)
{\it are
nonhomogeneous deformations
and curvature
of  congruence $\jmath_\lambda$}.
Another (equivalent) interpreting of nematic equilibrium equations
can by means of Kaluza-Klein-Maxwell equations:
\[\Rc_{\lambda h}=0 \quad \Leftrightarrow \quad
\varepsilon\dvg\mathcal{F}\jmath=
-\varepsilon {\rm grad}\,\overline{\mathcal{D}}+\varepsilon
{\rm div}\,\mathcal{D}\jmath+2\mathcal{F}\jmath(\jmath_\alpha,\ ).\]
Their combination gives the following, equivalent system of
nematic  equilibrium equations:
\begin{equation}\label{nemmax}
2\varepsilon\dvg\mathcal{F}\jmath=(3\mathcal{F}-\mathcal{D})\jmath(\jmath_\alpha,\  ).
\end{equation}
which does'nt contain derivatives of $\mathcal{D}.$

\section{Particular solutions} \label{part}

Let's consider some particular solutions to (\ref{nemmax}),
which corresponds to some earlier used  $\lambda.$

1. $\mathcal{F}=\mathcal{D}=0.$ This choice has been used
by a number of authors \cite{vlad,kok6}, who have investigated
4D physical properties of a
 "fifth coordinate independent" 5D physical world without electromagnetism.
An effective matter of the models is originated  only
from  $\jmath_\alpha,$ which in special coordinate system,
 adopted to $\jmath_\lambda,$ is proportional to ${\rm grad}\, \varphi,$
 where $\varphi=\sqrt{|G_{55}|}$ --- geometrical scalar field.

2. $\mathcal{F}=0\quad \jmath_\alpha=0.$
This choice has been used in works by a number of other authors
\cite{wes1}, where an effective matter involves
"fifth coordinate dependency" of 5D metric.
Easy to see, that, the canonical frame
of 5D metric $G=g_{\alpha\beta}(x,\eta)(dx^\alpha\otimes dx^\beta)
+\varepsilon d\eta\otimes d\eta,$
$\alpha,\beta=0,1,2,3,$ introduced in \cite{wes1} and in a number
of earlier works of the author,
just can be related to the considered particular
class of solutions to (\ref{nemmax}).

3. $\mathcal{F}=0;\ \mathcal{D}(\jmath_\alpha,\ )=0.$
This class is intermediate between  1 and 2.
It has'nt been investigated  in literature.

All considered cases imply $\mathcal{F}=0,$ in spite of the central
idea of KKT ---
geometrization of electromagnetic interactions. We'll discuss this circumstance
in Conclusion.

\section{Example: nematic structure of a flat 5D space-time}\label{ex}

Let $\mathcal{M}$ be flat 5D Minkowski space-time
with metric $G:$
\[
G=dt\otimes dt-dr\otimes dr-r^2d\varphi\otimes d\varphi-dz\otimes dz-d\eta\otimes d\eta,
\]
taken in 5D cylindrical coordinate system.
We'll treat $\mathcal{M}$ as infinite nematic medium with no boundaries.
Lets consider the following situations.

\begin{figure}[htb]
{\centering
\unitlength=0.7mm
\special{em:linewidth 0.4pt}
\linethickness{0.4pt}
\begin{picture}(137.00,46.00)
\bezier{160}(21.00,46.00)(39.50,44.50)(41.00,26.00)
\bezier{160}(1.00,26.00)(2.50,44.50)(21.00,46.00)
\bezier{160}(21.00,6.00)(2.50,7.50)(1.00,26.00)
\bezier{160}(41.00,26.00)(39.50,7.50)(21.00,6.00)
\bezier{80}(21.00,36.00)(30.25,35.25)(31.00,26.00)
\bezier{80}(11.00,26.00)(11.75,35.25)(21.00,36.00)
\bezier{80}(21.00,16.00)(11.75,16.75)(11.00,26.00)
\bezier{80}(31.00,26.00)(30.25,16.75)(21.00,16.00)
\bezier{120}(21.00,41.00)(34.87,39.87)(36.00,26.00)
\bezier{120}(6.00,26.00)(7.12,39.87)(21.00,41.00)
\bezier{120}(21.00,11.00)(7.12,12.12)(6.00,26.00)
\bezier{120}(36.00,26.00)(34.87,12.12)(21.00,11.00)
\emline{7.00}{12.00}{1}{35.28}{40.28}{2}
\emline{13.81}{33.15}{3}{15.72}{35.28}{4}
\emline{13.88}{33.15}{5}{16.38}{34.25}{6}
\emline{10.29}{36.75}{7}{12.20}{38.88}{8}
\emline{10.36}{36.75}{9}{12.86}{37.85}{10}
\emline{6.77}{40.34}{11}{8.68}{42.47}{12}
\emline{6.84}{40.34}{13}{9.34}{41.44}{14}
\put(21.00,26.00){\circle*{1.50}}
\put(35.00,41.00){\makebox(0,0)[lb]{$M_h(\varphi)$}}
\put(23.00,24.00){\makebox(0,0)[lt]{$P_3$}}
\put(10.00,32.00){\makebox(0,0)[cc]{$\jmath_\lambda$}}
\put(21.00,1.00){\makebox(0,0)[ct]{(1)}}
\bezier{160}(68.00,46.00)(86.50,44.50)(88.00,26.00)
\bezier{160}(48.00,26.00)(49.50,44.50)(68.00,46.00)
\bezier{160}(68.00,6.00)(49.50,7.50)(48.00,26.00)
\bezier{160}(88.00,26.00)(86.50,7.50)(68.00,6.00)
\put(68.00,1.00){\makebox(0,0)[ct]{(2)}}
\emline{48.00}{26.00}{15}{88.00}{26.00}{16}
\emline{57.77}{8.50}{17}{77.77}{43.14}{18}
\emline{57.83}{43.17}{19}{78.17}{8.67}{20}
\emline{62.90}{34.59}{21}{63.83}{31.91}{22}
\emline{62.90}{34.59}{23}{64.70}{32.51}{24}
\emline{72.92}{34.74}{25}{71.32}{33.11}{26}
\emline{72.92}{34.74}{27}{72.12}{32.42}{28}
\emline{78.00}{26.00}{29}{75.00}{26.67}{30}
\emline{75.00}{25.33}{31}{78.00}{26.00}{32}
\emline{58.00}{26.00}{33}{60.83}{26.67}{34}
\emline{60.83}{25.33}{35}{58.00}{26.00}{36}
\emline{62.86}{17.37}{37}{63.63}{19.98}{38}
\emline{62.86}{17.37}{39}{64.69}{19.32}{40}
\emline{73.08}{17.36}{41}{71.16}{19.15}{42}
\emline{73.08}{17.36}{43}{72.16}{19.94}{44}
\bezier{160}(115.00,46.00)(133.50,44.50)(135.00,26.00)
\bezier{160}(95.00,26.00)(96.50,44.50)(115.00,46.00)
\bezier{160}(115.00,6.00)(96.50,7.50)(95.00,26.00)
\bezier{160}(135.00,26.00)(133.50,7.50)(115.00,6.00)
\put(115.00,26.00){\circle*{1.50}}
\put(115.00,1.00){\makebox(0,0)[ct]{(3)}}
\emline{95.00}{26.00}{45}{135.00}{26.00}{46}
\put(115.00,26.00){\circle{4.00}}
\emline{135.00}{26.00}{47}{135.00}{38.00}{48}
\emline{95.00}{26.00}{49}{95.00}{14.00}{50}
\emline{95.00}{14.00}{51}{94.33}{17.33}{52}
\emline{95.67}{17.33}{53}{95.00}{14.00}{54}
\emline{135.00}{38.00}{55}{134.33}{34.67}{56}
\emline{135.67}{34.67}{57}{135.00}{38.00}{58}
\emline{100.00}{26.00}{59}{100.00}{19.00}{60}
\emline{109.00}{26.00}{61}{109.00}{24.00}{62}
\emline{130.00}{26.00}{63}{130.00}{33.00}{64}
\emline{125.00}{29.00}{65}{125.00}{26.00}{66}
\emline{100.01}{18.97}{67}{99.57}{21.28}{68}
\emline{100.44}{21.28}{69}{100.01}{19.05}{70}
\emline{104.99}{22.40}{71}{104.63}{23.79}{72}
\emline{105.34}{23.79}{73}{104.99}{22.44}{74}
\emline{104.99}{22.44}{75}{104.99}{25.98}{76}
\emline{109.01}{23.99}{77}{108.77}{25.03}{78}
\emline{109.25}{25.03}{79}{109.01}{23.99}{80}
\emline{120.98}{27.52}{81}{120.75}{26.67}{82}
\emline{121.20}{26.67}{83}{120.98}{27.48}{84}
\emline{120.98}{27.48}{85}{120.98}{26.00}{86}
\emline{124.98}{29.37}{87}{124.67}{28.06}{88}
\emline{125.30}{28.06}{89}{124.98}{29.32}{90}
\emline{130.02}{33.01}{91}{129.57}{30.67}{92}
\emline{130.47}{30.67}{93}{130.02}{33.01}{94}
\put(137.00,32.00){\makebox(0,0)[lc]{$\jmath_\lambda$}}
\put(68.00,34.00){\makebox(0,0)[cc]{$\jmath_\lambda$}}
\put(68.00,16.00){\makebox(0,0)[cc]{$P_3$}}
\bezier{20}(115.00,26.00)(127.00,28.00)(135.00,38.00)
\bezier{20}(115.00,26.00)(102.00,23.00)(95.00,14.00)
\put(82.00,41.00){\makebox(0,0)[lb]{$M_h(r)$}}
\end{picture}
\par}
 \caption{\label{fig}\small Nematic structures with cylindrical symmetry. In case (3) horisontal projection of field $\jmath_\lambda$ is shown. The vector field turns about radial direction and become vertical (orthogonal to a picture plane) at the center.}
\end{figure}
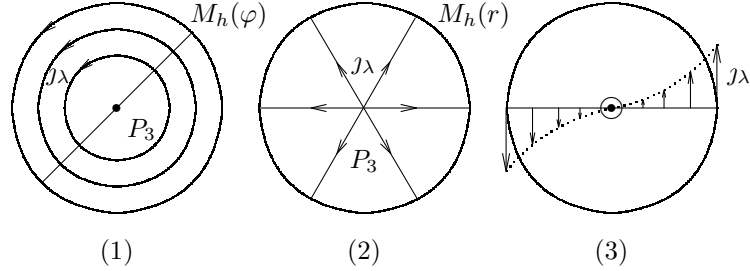

1. $\lambda=rd\varphi.$ Direct calculations gives:
\[
\alpha=d\ln r;\quad \mathcal{F}=\mathcal{D}=0,
\]
so equilibrium equations are satisfied identically (case 1).
4D space-times $\mathcal{M}_h(\varphi)$,
defined by $\lambda,$ are 4D pseudoeuclidian hyperplanes,
all going through  3-plane $\mathrm{P}_3:$ $r=0$ (see Fig.\ref{fig}(1)).

2. $\lambda=dr.$ Direct calculations give:
\[
\alpha=0;\quad \mathcal{F}=0;\quad \mathcal{D}=rd\varphi\otimes d\varphi,
\]
so equilibrium equations are satisfied identically (case 2).  Here physical worlds
$\mathcal{M}_h(r),$ defined by $\lambda,$ are 4D pseudoeuclidian coaxial cylinders (see Fig.\ref{fig}(2)).
$3-$plane
$r=0$ is   peculiar only in the sense that:
${\rm dim}\,\mathcal{M}_h(0)=3,$ rather then 4.

3. $\lambda=(1/\sqrt{\Delta})((r^2/r_0)d\varphi+dz),$
where $\Delta=1+(r/r_0)^2,$ $r_0={\rm const}.$
Direct calculations give:
\[
\alpha=\frac{r}{r^2_0\Delta}dr;\quad \mathcal{F}=
\frac{r}{r_0\Delta^{3/2}}(dr\wedge d\varphi-\frac{1}{r_0}dr\wedge dz);\quad
\mathcal{D}=0.
\]
This solution possesses nonzero nematic energy density and corresponds to
the solution\footnote{The solution, obtained in this book is valid
under $K_3>K_2.$ Our solution ($K_3=0$) is not contained in those,
performed on p.203 of \cite{land}.} of problem 1 in  $\S38$ of \cite{land}.
Total 5D nematic energy of configuration (per element of infinite 3D volume),
is
\[
\mathcal{E}=\frac{K_2}{12}\int\limits_{R^2} r\,dr\, d\varphi(\lambda\wedge d\lambda)^2
=-2\pi K_2.
\]
Under $r\to\infty$ $\lambda_\infty=r d\varphi,$ under
$r\to 0$ $\lambda_0=dz.$
In Fig.\ref{fig}(3) nematic structure of the solution is performed.

Examples 1 and 2 perform nematic structures with linear topological defects,
which are called {\it disclinations.} We'll discuss them in Conclusion.

\section{Conclusion}

{\bf 1. Nature of 5D nematic structure.} Nematic structure of common fluid crystals
is originated from special kind of interaction between individual molecular dipoles.
Generic thread-like nematic structure  provides total dipole-dipole interaction energy minimum by means of more or less
clear 3D physical mechanism.

In case of 5D KKT there is no such clear physical mechanism of how individual vectors
$\jmath_\lambda(p)$ form smooth vector field $\jmath_\lambda$ on $\mathcal{M}.$
Moreover, we don't know even what is physical nature of an individual vector
$\jmath_\lambda(p),$ attached to  every point $p\in\mathcal{M}.$
However, the role of $\jmath_\lambda$ in KKT --- extracting of 4D {\it observable}
physics from 5D geometry,
suggests, that nature of $\jmath_\lambda$ should concern some subtle
aspects of relations between observer consciousness and multidimensional
physical (or more exactly geometrical) reality. This view adjoins Penrose's
"physics of consciousness" (PhC) \cite{penrose} in the multidimensional physics
context. Present work should be regarded as "macroscopic" (or "phenomenological")
model of PhC, which don't touch origins and internal structure\footnote{Similarly
to statistical physics of liquid crystals, $\lambda$ could be viewed as some
collective property of "mental molecules." In this context
monad field $\lambda$ resembles Leibniz's metaphysical "monad" not only
terminologically.}
of $\lambda.$ But even this "averaged" approach, inspired by 5D KKT,
suggests the following conclusion: {\it  it would be no 4D space-time and matter
without nematic structure of 5D space-time.}
When 5D nematic structure is  destroyed
(due to some extreme conditions, originated from a more
rich 5D geometrical  model, including nonlinearity and thermodynamics),
4D space-time and matter disappear and we have
chaotically distributed horisontal tangent spaces in $\mathcal{M}$
or "space-time-matter chaos."

{\bf 2. Boundary conditions.} In common  nematic crystals physics boundary
interactions of nematic molecules as a rule are much more intensive then volume
ones. It allows operate nematic structure   in laboratory experiments.
From the theoretical viewpoint boundary conditions fix integration constants
in general solutions to equilibrium equations.

In case of 5D nematic, viewed as object of PhC, status of 5D boundary
conditions is unclear. Probably, we should reverse our consideration:
fixing integration constants in general solution by comparing 4D physics,
involved by $\lambda$, with observable 4D world,
one could try to determine (may be under some additional assumptions)
5D boundary conditions (see discussion in \cite{kok8}).

{\bf 3. Interaction with curvature and others fields.}
In present work we have restricted ourself by the simplest 5D model ---
pure geometry of vacuum 5D space-time. It has led us to a rather "poor"
physics, involving only twist deformations of 5D nematic media.
Contrary to this 5D model, common 3D nematics,
possessing all Frank's moduluses, are able to form rich topological structures due to
a subtle balance of splay, twist and bend energies.  Moreover, external
electromagnetic fields can operate director's field inside nematic sample.
This property is widely used in fluid crystal screens (Freedericksz's effect).

The similar situation can be obtained in 5D nematic
by suitable generalization of our model. 5D Ricci curvature, nonlinearity
or nonriemannian objects -- torsion and nonmetricity -- will force
5D nematic structure as some "external fields." We put off all this possibility
for a future work.

{\bf 4. Physics of nematic topological defects.} Let's turn our attention to
example 1 of Sec.\ref{ex} with circular nematic structure. In spite of flatness
4D worlds $\mathcal{M}_h(\varphi)$ possess  the remarkable property: {\it since
$\lambda|_{\mathrm{P}_3}=0,$ then
$\ker\lambda|_{\mathrm{P}_3}=(T\mathcal{M})|_{\mathrm{P}_3},$ i.e. tangent  to
peculiar 3D pseudoeuclidian plane spaces are 5D, rather then 4D (or 3D).}
What  are  physical manifestation of such peculiarity?

Let some test particle moves along 5D geodesic $\gamma(\tau)$ with 5D
velocity $U,$ satisfying 5D geodesic equations: $\nabla_UU=0.$
When such geodesic will be horizontal (i.e. $d\gamma/d\tau\in T_h\mathcal{M}$
for every $\tau\in R$)? In case $U_\lambda=\dot U_\lambda=0$ at some point of $\gamma,$
decomposition (\ref{deccov}) gives at the same point:
\[
\nabla_UU=\frd\nabla_{U_h}U_h-\varepsilon\mathcal{D}(U_h,U_h)\jmath_\lambda
\]
or equivalently
\[
\nabla_UU=0|_{U_\lambda=0}\Leftrightarrow\frd\nabla_{U_h}U_h=0;\quad
\mathcal{D}(U_h,U_h)=0.
\]
We see, that free test particle, moving at  some point with horizontal velocity
$U_h$, will continue to move along 4D geodesic,
if $\mathcal{D}(U_h,U_h)=0$ at any $\tau\in R.$
In our example $\mathcal{D}=0,$ so any free test particle, living on some
regular part of $\mathcal{M}_h(\varphi),$ will move there rectilinearly,
always being attached  to the $\mathcal{M}_h(\varphi).$

This picture is violated on $\mathrm{P}_3.$ Here horizontal tangent space, spanned by
all possible directions of initial velocities, is whole $T\mathcal{M}|_{\mathrm{P}_3}$
(or, more exactly, interior of 5D light cone).
It means, that {\it $\mathrm{P}_3$ is 3D region of 4D worlds, where one can send
particles and signals in extradimension or receive them from there.}
In other words, $\mathrm{P}_3$ could be looked as the place,
where 4D conservation laws are violated (while 5D ones are, of course, valid).

Another peculiar property of $\mathrm{P}_3$ follows from the relation:
\[
\mathrm{P}_3=\bigcap\limits_{\varphi=0}^{2\pi}\mathcal{M}_h(\varphi).
\]
So, $\mathrm{P}_3$ also can be viewed as "junction station"
for travels\footnote{Note, that traveling objects should be matter
points, lines or planes, since $\mathrm{P}_3$ is 3D pseudoeuclidian space
with 2D space section.}
from one $\mathcal{M}_h(\varphi_1)$ to another
$\mathcal{M}_h(\varphi_2).$

{\bf 5. Topological classification   of defects and Frank's indexes.}
Circular and radial defects, shown in Fig.\ref{fig}, are particular cases
of nematic disclinations, possessing cylindrical symmetry. Simple physical
considerations (radial self-similarity and uniqueness, see \cite[\S39]{land})
show, that any cylindrical disclination an be described by a winding
number $n=0,\pm1/2,\pm1,\pm3/2\dots,$ which is equal to a number of director revolutions
under moving along closed path, embracing defect line.
In case of cylindrical symmetry $n$ is called {\it Frank's index} of topological
defect. Vicinity of a defect with Frank's index $n$ possesses axe of symmetry
$D_m,$ where $m=2|n-1|,$  which, besides rotations by angles $\varphi=2\pi p/m,
p=0,\dots,m,$ includes reflections with respect to horizontal plane, orthogonal to
axe of the defect. Note, that the defects, considered in examples 1 and  2
both have Frank's index  $n=1.$ More deep topological analysis  reveals,
that all defects with integer $n$ can be eliminated by continuous deformation
of director field, while disclinations with half-integer $n$ are topologically
inherent. Mathematically this facts are originated from a structure of fundamental group
of nematic configuration space $RP^2.$ As it can be shown by algebraic topology
methods \cite{boris}, the fundamental group $\pi_1(RP^2)=\mathbb{Z}_2.$
In other words, all closed contours on $RP^2$ are divided  on the two
subclasses:
closed and "semiclosed", with end points lying on diameter of projective
sphere $RP^2.$ Integer Frank's indexes correspond to the first class, half-integer --
to the second. Physially, inherent disclinations are topologically stable and can be
observed in laboratory, while eliminable ones are destroyed by small
external uncontrolled influences. Point-like defects of 3D nematic can
be also topologically classified by means of structure of second fundamental group
$\pi_2(RP^2)=\mathbb{Z}.$ It turns out, that point-like defect is stable if
its topological number (named sometimes "topological charge") $n\neq 0.$

In spite of more possibilities of $n-$dimensional nematic structures,
topologically they copy 3D case. Really, configuration space of
$n-$dimensional nematic  is $RP^{n-1}.$  Topological classification of
$k-$dimensional defects $(0\le k\le n-2)$ will be established on structure
of fundamental group $\pi_{n-k-1}(RP^{n-1}).$ But all this group are well known:
\[
\pi_{n-k-1}(RP^{n-1})=\left\{ \begin{array}{lr}
\mathbb{Z},&k=0;\\
0,& 0<k<n-2;\\
\mathbb{Z}_2,& k=n-2.
\end{array}
\right.
\]
So, the case $k=0$  is topological analog of  point-like defects of
 3D nematic, while multidimensional topological analog of line defect in 3D
 is the case $k=n-2.$ All defects of intermediate dimensions
 are topologically trivial.

In a difference with topology, group theoretical structure
of symmetry of multidimensional defect vicinity will be really
more rich, then in case of 3D. The similar to 3D case considerations
(in flat $n-$dimensional  Minkowski space) lead to conclusion, that
director field  in vicinity of defect (of any dimension $0\le k\le n-2$)
will have symmetry of some discrete subgroup $\mathcal{S}\subset O(1,n-1),$
whose elements (their number and type) can be viewed as
"generalized Frank's indexes" of the multidimensional nematic  defects.

{\bf 8. Physics on anholonomic manifold.}
Due to the fact, that KKT consider twist tensor $\mathcal{F}$ both geometrically --
as object, responsible for anholonomicity of horizontal 4D space-time, and physically --
as (related to) geometrized electromagnetic strength tensor field,
{\it Kaluza-Klein electrodynamics is not identical to standard Maxwell one.}
This fact had led authors, who had worked with KKT, to a number
of  "fine tunings" of the theory, which had made equations
of Kaluza-Klein electrodynamics compatible with Maxwell equations.
For example, general $\mathcal{F}$ is not suitable candidate for direct
geometrization of electromagnetic field, since $d\mathcal{F}\neq0.$ One of the ways to obtain
standard "second pair" of Maxwell equations is to specialize some
geometrical objects. Let $\alpha$ be exact form, i.e. $\alpha=d\ln\psi$ and
let\footnote{Opposite sign $\varepsilon=-1,$ accepted in \cite{vlad}
is originated from the opposite sign on definition of curvature.}
$\varepsilon=+1.$ Then, assuming electromagnetic potential form
$A=(1/2)\sqrt{\kappa_5/8\pi l_5}\psi\lambda,$ where $l_5$ --- "size of $\mathcal{M}$
in fifth dimension", we obtain:
\[
dA=(1/2)\sqrt{\kappa_5/8\pi l_5}(d\psi\wedge\lambda+\psi d\lambda)=
\sqrt{\kappa_5/8\pi l_5}\psi\mathcal{F}.
\]
Then after $4D$ conformal transformation $h=\psi^{-1}\tilde h,$ where
$\tilde h$ --- physical (observable) 4D metric, we formally
obtain from (\ref{act4}) "right geometrized action"   for electromagnetic field:
\begin{equation}\label{actem}
\fvd\mathcal{A}_{\rm em}=-\frac{1}{16\pi}\int\limits_{\mathcal{M}}
(dA)^2\, \vl{5}.
\end{equation}
But if $\lambda\wedge d\lambda=2\lambda\wedge\mathcal{F}\neq0,$
i.e. (after projection on $\jmath_\lambda$) $\mathcal{F}\neq0,$
then
\[
\mathcal{M}\neq\bigcup\limits_{\eta\in R}\mathcal{M}_h(\eta),
\]
where $\mathcal{M}_h(\eta)$ --- classical submanifolds of $\mathcal{M},$
i.e. {\it separate integration over extradimension  in (\ref{actem})
is impossible.} Formal decomposition $\vl{5}=\lambda\wedge\vl{4}^h$
and integration over $\mathcal{M}_h$ with volume form $\vl{4}^h$
will give infinite integrals, since by Rashewski-Chow's theorem
anholonomic manifold $\mathcal{M}_h$ is 5D (as a set) and it will have
infinite measure with respect to 4D volume form $\vl{4}^h.$
Roughly speaking, {\it within 5D KKT we are able to derive  4D
differential (local) laws of physics, induced by 5D geometry,
while integral laws, generally speaking, are absent (or should be modified).}
So, standard Gauss theorem and Coulomb's law will be different from those,
inspired by 5D KKT, since  sphere, which is commonly used
in derivation of Coulomb's law
from Maxwell equations,  on anholonomic manifold  has
nothing to do
with common sphere (see \cite{griffits}).

\end{document}